\documentstyle[prd,preprint,aps]{revtex}
\begin{document}
\tighten
\draft
\preprint{ }
\title{Response of Mica to Weakly Interacting Massive Particles}
\author{J. Engel}
\address{Department of Physics and Astronomy,
University of North Carolina, Chapel Hill\\ North Carolina 27599  }
\author{M.T. Ressell}
\address{W.K. Kellogg Radiation Laboratory, 106-38, Calfornia Institute
of Technology, Pasadena, CA 91125}
\author{I.S. Towner}
\address{AECL, Chalk River Laboratories, Chalk River, Ontario, Canada
K0J 1J0}
\author{W.E. Ormand}
\address{Physics Department, 401 Nielsen Hall, University of Tennessee,\\
Knoxville, TN 37996-1200\\
 and \\
Physics Division, Oak Ridge National Laboratory, P.O. Box 2008, \\
MS-6373 Building 6003, Oak Ridge, TN 37831-6373}
\date{\today}
\maketitle
\begin{abstract}

We calculate spin-dependent cross sections for the scattering from mica of
hypothetical weakly interacting dark-matter particles such as neutralinos.  The
most abundant odd-A isotopes in mica, $^{27}$Al and $^{39}$K, require
different shell-model treatments.  The calculated cross sections will allow the
interpretation of
ongoing experiments that look for tracks due to the interaction of
dark-matter particles with nuclei in ancient mica.

\end{abstract}

\narrowtext

\section{Introduction}
\label{s:intro}

The nature of the dark matter in our galaxy and elsewhere has
become increasingly puzzling\cite{r:DM1}.  Although it is too early to
make definitive statements, ongoing experiments seem to imply
that there are not enough macroscopic objects in the galactic
halo\cite{r:machos} to account for the gravitational attraction felt by
luminous objects near the edge of the disk.  One alternative is
dark matter in the form of elementary particles. For several years now,
weakly-interacting massive particles (WIMPs) that arise in
supersymmetric extensions of the standard model have been attractive
candidates\cite{r:DM2}.  A variety of experiments to detect WIMPs are
either already operating or in the planning/prototype stage.

One promising project\cite{r:Price} uses ancient mica as a detector.
The idea is that over long periods of time a countable number of WIMPs should
have collided with underground mica and left recognizable tracks.  At present,
the experiment is not sensitive enough to test the WIMP hypothesis fully, but
by increasing the amount of mica examined, experimenters hope to reach the
required level of sensitivity within a few years.  Then,
confirming or rejecting the WIMP hypothesis
will require a knowledge of their cross sections for scattering from the
various
elements in mica.  In this paper we calculate these cross sections so that the
experiments can be properly interpreted.  Our results are general enough to be
applied to any heavy Majorana particle; we explicitly calculate the
``spin-dependent" response of mica.  While ``spin-independent" cross sections
for supersymmetric neutralinos\cite{r:Drees}, perhaps the most plausibly
motivated WIMPs,
are usually larger than the spin-dependent cross sections in most of
the nuclei in mica, there are still regions of
supersymmetric parameter space for
which this is not so.  Furthermore, WIMPs with no scalar interactions, such as
heavy Majorana neutrinos, have not yet been completely ruled out.  A good
calculation of the spin-dependent response is therefore warranted.  The
spin-independent scattering can be easily calculated following, for example,
the work
of Ref.\ \onlinecite{r:Engel}.

The composition of mica is 58\% $^{16}$O, 16\% Si (mostly $^{28}$Si), 12\%
$^{27}$Al, 5\% K (mostly $^{39}$K), and 9\% H.  Since the scattering
from even-even
nuclei is entirely spin-independent --- and therefore easily calculable --- we
need to concentrate here only on $^{27}$Al and $^{39}$K.  (No detailed
nuclear-structure calculation is required for H.)
The nucleus $^{27}$Al
is also one of the active ingredients in a very high-resolution ($\Delta
E_{FWHM} \approx 200$ eV) and low-threshold ($E_{th} \approx 500$ eV)
sapphire-crystal (Al$_2$O$_3$) detector that is currently under
development\cite{r:Cooper}.  The low threshold of this experiment makes it
ideal for detecting lighter WIMPs.

Though both aluminum and potassium nuclei are nominally classified to be
in the $sd$ shell, $^{39}$K is at the upper edge of the shell and its wave
functions contain significant admixtures of 1- and 2- particle-hole states
involving higher shells.  Most of the correlations in $^{27}$Al, by contrast,
are produced by purely $sd$-shell configurations.  The techniques we apply in
the two cases are therefore quite different.  After a general discussion of
spin-dependent scattering in Section \ref{s:general}, we will examine the two
odd-A nuclei independently in Sections \ref{s:Al} and \ref{s:K}.  In section
\ref{s:d} we briefly discuss the implications of these calculations for
mica-based detection.

\section{Spin-Dependent Scattering}
\label{s:general}

At low-momentum transfer, the tree-level coupling of neutralinos to quarks
(excluding the exchange of Higgs bosons) yields the elastic scattering
amplitude from a nucleus $|N \rangle$:
\begin{equation}
\label{e:amp}
M = A \langle N | a_p {\bf S}_p + a_n {\bf S}_n | N \rangle \cdot
{\bf s}_{\chi} ~~,
\end{equation}
where ${\bf S}_p$ and ${\bf S}_n$ are the total proton and neutron spins in the
nucleus, ${\bf s}_{\chi}$ is the spin of the neutralino, $a_p$ and $a_n$ are
coupling constants that depend on the composition of the neutralino (in terms
of
the four or more neutral fermionic partners of the gauge and Higgs bosons) and
on the
distribution of nucleon spin among the different types of quarks, and the
quantity $A$ depends on other physics at the supersymmetry
scale\cite{r:Griest}.
Though these specifics apply only to neutralinos, the spin-dependent scattering
amplitudes for other Majorana WIMPs will have the same general form, Eq.
(\ref{e:amp}).  The spin-dependent cross section is a product of the square of
the amplitude and a phase-space factor.

When the momentum transferred to the nucleus by the WIMP, $q$,
becomes comparable to
the inverse radius of the nucleus,
structure functions (form factors) resembling
those associated with elastic magnetic electron or neutrino scattering modify
Eq.\ (\ref{e:amp}).  The more general expression for the spin-dependent cross
section is\cite{r:Engel}
\begin{equation}
{d \sigma \over dq^2} = {8G_F^2 \over (2J+1) v^2} S_A(q)~,
\end{equation}
where $G_F$ is the Fermi weak four-point effective coupling constant,
and $v$ the neutralino velocity.
The subscript $A$ stands for ``axial". The function $S_A(q)$ can always be
written in terms of three fundamental structure functions
\begin{equation}
S_A(q) = a_0^2 S_{00}(q) + a_1^2 S_{11}(q) + a_0 a_1 S_{01}(q)~.
\end{equation}
The labels 0 and 1 refer to isospin; the constant $a_0$ is given by $a_p + a_n$
and $a_1$ by $a_p - a_n$.  The three structure functions
contain expectation values of operators of the form
$j_L(qr) [ Y_L \sigma]^{L \pm
1}$ ($L$ even), which depend on spatial coordinates as well as nucleon spins.
Details appear in Ref.\ \onlinecite{r:Engel}.
Without rewriting all the expressions here, we describe
the calculation of $S_A(q)$ in
$^{27}$Al and $^{39}$K.

\section{Aluminium}
\label{s:Al}

The nucleus $^{27}$Al lies in the middle of the $sd$ shell.
To obtain the ground-state wave
function we diagonalize Wildenthal's ``universal" $sd$ interaction
\cite{r:Wildenthal} in the full 0$\hbar \omega$ space.  This
interaction has been meticulously developed and tested over many years and
together with effective operators accurately reproduces most low-energy
observables in {\it sd}-shell nuclei\cite{r:Brown1}.  We carry out our
calculations with the Lanczos m-scheme shell-model code
CRUNCHER \cite{r:Resler}
and its auxiliary codes; the m-scheme basis for $^{27}$Al contains 80115 Slater
determinants.  A very similar calculation for $^{29}$Si is
reported in \cite{r:Ressell}, where more
details appear.

The agreement of the calculated spectrum with that
measured in $^{27}$Al is excellent,
reflecting the effort that went into constructing the interaction.  More
relevant is the accuracy
of the calculated magnetic moment, which derives from the
same spin operators that
determine the WIMP structure functions at $q^2 = 0$.  The measured magnetic
moment is $\mu_{exp} = +3.6415~\mu_N$.  Using free-particle
$g$-factors and our wave functions, we obtain a value
$\mu_{calc} = +3.584~\mu_N$,
which agrees well with experiment.  By contrast, the single-particle
value is $\mu_{sp} = +4.793~\mu_N$.

Although Ref.\ \onlinecite{r:Brown1} advocates
the use of effective operators in the $sd$ shell, we obtain a calculated
magnetic moment in $^{27}$Al closer to experiment
with the free-nucleon spin and orbital-angular-momentum
operators (the same was true in $^{29}$Si).  Unfortunately, we
overpredict the
Gamow-Teller matrix element by a factor of about 1.25,
a problem that the effective
operators would remedy.  Thus,
we have conflicting evidence on whether to use effective operators or not.
In the absence of a better prescription, we choose to use the
free-nucleon g-factors.
The consequence for WIMP scattering is that we did not quench
our calculated cross sections at all.  In $^{39}K$, where our calculation
provides more information, we make a similar decision.

Ref.\ \onlinecite{r:EV} develops two schemes, based on the ``Odd Group Model"
(OGM) and an extension of it (EOGM) for extracting values of neutron and proton
spins --- the quantities that determine spin-dependent WIMP cross sections at
$q^2 = 0$ --- from magnetic moments and beta-decay lifetimes.  These values for
$^{27}$Al appear in Table \ref{t:1} alongside the spins and orbital angular
momenta from our calculation.  The most surprising result is the large
value for $\langle {\bf L}_n \rangle$ obtained in the shell-model calculation.
On comparing the numbers in Table
\ref{t:1}, we see
that the single-particle model (ISPSM) and OGM disagree with the shell model;
the large value of the neutron's
orbital angular momentum is the
likely explanation of the OGM's failure to reproduce the spin angular momenta.
By contrast, the EOGM results
with $g_A/g_V = 1$ (a quenching of the Gamow-Teller matrix element) are very
close to our shell-model values.  This agreement with the most sophisticated
phenomenological analysis is heartening. However,
since spin-dependent neutralino scattering involves the axial nuclear current,
the failure of any analysis to reproduce beta-decay lifetimes accurately
without ad hoc quenching --- a problem not mentioned in any previous
work --- is troubling.  As a result our aluminum calculation has an
uncertainty from this source of roughly 30\%.  The situation is better in the
potassium calculation since the core-polarization corrections,
to be discussed in the next section, explain much of
the beta-decay quenching.

As described above, when heavy particles transfer momentum comparable to
the inverse nuclear radius, structure functions modify the simple form Eq.\
(\ref{e:amp}) for the scattering amplitude.  In the shell model, the
expectation value of any one-body operator is easy to calculate,
so the evaluation of structure functions poses no additional problem.
We did, however,
improve upon an approximation in previous works, namely the use of
harmonic oscillator single-particle wave functions;
the true nuclear mean-field potential
is closer to a Woods-Saxon than a harmonic oscillator form.
For comparison we
calculate the structure functions with both types of wave functions.
We use a length parameter of $b=1.73$ fm for oscillator functions  and
the standard parameters of Ref.\ \onlinecite{r:Donnelly} for the Woods-Saxon
potential.  Fig.~\ref{f:1} shows the three structure functions
in both cases as a
function of $q^2$ up to the maximum possible momentum
transfer for a WIMP moving at 700 km/sec (more than twice
the mean WIMP velocity).
The Woods-Saxon wave functions make only small changes in the results.

For the momentum transfers in the figure, the Woods-Saxon structure functions
are
given to high accuracy by the third-order polynomials
\begin{eqnarray}
\label{e:WS}
S_{00} &  = & 0.0929516 - 0.472059\, y + 1.05996\, y^2 -1.01148\, y^3   \\
\nonumber
S_{11} & = & 0.0657232 - 0.449840\, y + 1.35041\, y^2 -1.68508\, y^3  \\
S_{01} & = & 0.1563300 - 0.935958\, y + 2.45779\, y^2 -2.72621\, y^3 ~,
\nonumber
\end{eqnarray}
where $y=(bq/2)^2$.
As noted above, these three functions allow the calculation of spin-dependent
cross sections for any heavy Majorana particle.

\section{Potassium}
\label{s:K}

For the nucleus $^{39}$K, a shell-model diagonalization is difficult to perform
because of the severe truncations required to the active model space.  Since
this nucleus is so near the boundary between the {\it sd} and {\it pf} shells,
excitations of particles into higher shells can have significant
effects that are often not well simulated by effective operators.  Despite the
difficulties, our first attempt to treat this nucleus was within the framework
of the shell-model, with the Hamiltonian and method advocated in Ref.\
\onlinecite{r:WBB}.  Computational complexity, however, kept us from allowing
more than two particles out of the {\it sd} shell into the {\it fp} shell.
After diagonalization, the 2p-3h excitations we did include made up
approximately 40\% of the ground-state wave function.  While the excitations
changed the orbital angular momentum significantly, they had almost no effect
on the spin, and our magnetic moment disagreed badly with experiment,
indicating that further correlations are indeed important.  Any further
complexity makes use of the shell model
difficult, however.  We turn instead to an alternative scheme, based
on perturbation
theory, that was successfully implemented a few years ago in calculations of
several spin-dependent observables in closed-shell-plus (or minus)-one
nuclei\cite{r:To87}.  Since this method has never before been applied to WIMP
structure functions, we describe it in more detail than we did our $sd$
shell-model results.

We begin by dividing the Hamiltonian into a one-body term and a
residual interaction:  $H = H_0 + {\cal V}$, where $H_0 = T + U$, the sum of
kinetic energy and one-body potential energy operators, and ${\cal V} = V - U$,
the difference between the two- and one-body potential energy operators.  The
eigenfunctions of
$H_0$, which we take to be a harmonic oscillator, form the basis of the
calculation.  The unperturbed ground state of $^{39}$K, a
single-proton hole in the $d_{3/2}$ orbital, gives the single-particle
(Schmidt) magnetic moment: -0.676 $\mu_N$.  Corrections come from the
perturbative expansion
in the residual interaction.  Very generally, to first order in a
closed-shell-plus (or minus)-one nucleus the matrix element of a one-body
operator, $F$, is given by
\begin{equation}
\label{firsto}
\langle \psi_b \! \mid \! F \! \mid \! \psi_a \rangle =
\langle b \! \mid \! F \! \mid \! a \rangle  +
\sum_{\alpha \neq a,b} \frac{\langle b \! \mid \! F \! \mid \! \alpha \rangle
\langle \alpha \! \mid \! {\cal V} \! \mid \! a \rangle}{E_a - E_{\alpha}} +
\sum_{\alpha \neq a,b} \frac{\langle b \! \mid \! {\cal V} \! \mid \! \alpha
\rangle
\langle \alpha \! \mid \! F \! \mid \! a \rangle}{E_b - E_{\alpha}}
\end{equation}
Here $a$ and $b$ are single-hole valence states and $\alpha$
an infinite set of single-hole or two-hole one-particle (2h-1p)
states constructed from the eigenfunctions of $H_0$; in our work
$a = b = 0d_{3/2}$.
The energy denominators also come from $H_0$; they are integral
multiples of $2 \hbar \omega$, where $\hbar \omega$ is the
characteristic oscillator energy.  We choose $\hbar \omega =
45 A^{-1/3} -25 A^{-2/3} = 11.1 \, {\rm MeV}$, corresponding to $b = 1.934$ fm,
for $A = 39$.  We explicitly evaluate the intermediate-state summations
for denominators equal to $2 \hbar \omega$,
$4 \hbar \omega$,
$6 \hbar \omega$,
and $8 \hbar \omega$, and then extrapolate the results geometrically.

The two-body matrix elements of ${\cal V}$ in eq.\,(\ref{firsto}) contain the
one-body potential through the $-U$ components, known as Hartree-Fock
insertions\cite{r:EM71}.  If the unperturbed Hamiltonian $H_0$ were chosen
to be the Hartree-Fock Hamiltonian that minimized the energy of a single Slater
determinant characterizing the closed-shell core, then the Hartree-Fock
insertions would have no effect.  Ref.\ \onlinecite{r:To87} shows that even
with an oscillator Hamiltonian the Hartree-Fock insertions do not affect
magnetic moments and their effects ought to be small for the operators
discussed here.  We therefore drop the Hartree-Fock insertions altogether.
Furthermore, all first-order corrections to magnetic moments, and to the
operators used here in the $q=0$ limit,
also vanish because the spin and orbital-angular-momentum
operators cannot create (or annihilate) a particle-hole state at $LS$
closed shells (selection rules require the particle and hole states to have the
same orbital structure).  At non-zero $q$, the first-order corrections to
$S_A(q)$, while not identically zero, are still small.  It is essential,
therefore, to
consider perturbation theory through to second order, and we add the following
additional terms to Eq.\ (\ref{firsto}):
\begin{eqnarray}
\label{secondo}
& & \sum_{\alpha , \beta \neq a,b} \langle b \! \mid \! F \! \mid \! \alpha
\rangle
\frac{ \langle \alpha \! \mid \! V \! \mid \! \beta \rangle \langle \beta \!
\mid \!
V \! \mid \! a \rangle }{(E_a - E_{\alpha})(E_a - E_{\beta}) }
- \sum_{\alpha \neq a,b } \langle b \! \mid \! F \! \mid \! \alpha \rangle
\frac{ \langle \alpha \! \mid \! V \! \mid \! a \rangle \langle a \! \mid \!
V \! \mid \! a \rangle }{ (E_a - E_{\alpha})^2 }
\nonumber  \\[1mm]
& & + \sum_{\alpha , \beta \neq a,b} \frac{ \langle b \! \mid \! V \! \mid \!
\beta \rangle \langle \beta \! \mid \! V \! \mid \! \alpha \rangle }
{(E_b - E_{\beta})(E_b - E_{\alpha}) } \langle \alpha \! \mid \! F \! \mid \!
a \rangle - \sum_{\alpha \neq a,b} \frac{ \langle b \! \mid \! V \! \mid \! b
\rangle \langle b \! \mid \! V \! \mid \! \alpha \rangle }
{(E_b - E_{\alpha})^2} \langle \alpha \! \mid \! F \! \mid \! a \rangle
 \\[1mm]
& & + \sum_{\beta , \gamma \neq a,b} \frac{ \langle b \! \mid \! V \! \mid \!
\beta \rangle }{(E_b - E_{\beta})} \langle \beta \! \mid \! F \! \mid \! \gamma
\rangle \frac{\langle \gamma \! \mid \! V \! \mid \! a \rangle }{(E_a - E_{
\gamma})} - \frac{1}{2} \sum_{\beta \neq a,b} \frac{ \langle b \! \mid \!
V \! \mid \! \beta \rangle \langle \beta \! \mid \! V \! \mid \! b \rangle}
{(E_b - E_{
\beta})^2} \langle b \! \mid \! F \! \mid \! a \rangle
\nonumber  \\[1mm]
& & - \frac{1}{2} \sum_{\beta \neq a,b} \langle b \! \mid \! F \! \mid \! a
\rangle
\frac{ \langle a \! \mid \! V \! \mid \! \beta \rangle \langle \beta \! \mid \!
V \! \mid \! a \rangle }{(E_a - E_{\beta})^2} ~.\nonumber
\end{eqnarray}
Here, as before, $a$ and $b$ are single-hole valence states,
$\alpha$ an infinite set of 2h-1p states, and $\beta , \gamma$ an
infinite set of 2h-1p and 3h-2p states.  Again the selection rules on
$\langle b \! \mid \! F \! \mid \! \alpha \rangle $ and $\langle \alpha \! \mid
\! F
\! \mid \! a \rangle $ forbid the first four terms of eq.\,(\ref{secondo})
from contributing at $q = 0$, and only allow small contributions at
higher $q$.  The fifth, sixth and seventh terms, however, each
contain intermediate-state summations over 2h-1p and 3h-2p
states that are not constrained by selection rules and converge only slowly
with increasing energy denominators.
[The last two terms renormalize the single-hole matrix element $\langle b
\! \mid \! F \! \mid \! a \rangle$.]  These three terms are sometimes called
the number-conserving set\cite{r:ES71} because if the one-body operator were
replaced by the number operator the terms would contribute nothing.  This is
apparent when the last three terms in Eq.\ (\ref{secondo}) are rewritten
in the following equivalent form (with $a=b$):
\begin{equation}
\label{commu}
\left \langle a \left | V \frac{Q}{e} \left [ F, \frac{Q}{e} V
\right ] \right | a \right \rangle ~,
\end{equation}
where the operator $Q/e$ is $\sum_{\beta} \mid
\beta \rangle \langle \beta \mid /(E_a - E_{\beta})$.
If the one-body operator $F$ were to commute with both the
energy denominator $e$ and the residual interaction $V$, as does the
number operator, the second-order correction to the diagonal matrix element
would vanish.  But spin-dependent one-body operators do not commute with
spin-dependent residual interactions, resulting, for example, in
second-order corrections to magnetic moments from tensor forces\cite{r:SIA74}.

Here we try two different residual interactions.  The first
(which we call $I$),
used in Ref.\ \onlinecite{r:To87} for magnetic-moment calculations, is
based on the one-boson-exchange potential of the Bonn type\cite{r:Bonn87}, but
limited to the four or five important meson exchanges.  For use in finite
nuclei, this interaction should be converted to a G-matrix; here we approximate
the procedure crudely by introducing a short-range correlation function.  The
resulting interaction has a weak tensor-force component typical of
Bonn potentials.  Our second interaction\cite{r:HKT85} (called $II$ here) is a
full G-matrix constructed from the Paris potential\cite{r:Paris80} and
parameterized in terms of sums over Yukawa functions of various ranges and
strengths.  Interaction II has a strong tensor force, producing
effects described below.

Our results for the orbital angular momenta and spins in $^{39}$K appear in
Table \ref{t:2}.  As was the case earlier, the spins from both forces agree
well with those obtained in the phenomenological EOGM.  To test our results
against data, we perform a comprehensive calculation of magnetic moments and
Gamow-Teller matrix elements, including (under the heading $MEC$)
meson-exchange currents, isobar
currents, and other relativistic effects (see\cite{r:To87}) in addition to the
core polarization described above.  Table \ref{t:3} presents the results for
both isoscalar and isovector quantities [The magnetic moment of $^{39}$K is the
sum of the isoscalar and isovector moments.]  The $MEC$ correction is small for
isoscalar magnetic moments and beta decay, but is significant for isovector
electromagnetic operators.  There is little difference between the two
interactions for isoscalar operators and the results are in good agreement with
experimental values.  For the isovector operators, interaction II produces
larger corrections and poorer agreement with the experimental isovector
magnetic moment.  But, although neither interaction reproduces the Gamow-Teller
beta-decay matrix element well, interaction II does better, and we
recommend it for that reason.  Compared with ${}^{27}$Al, much smaller ad hoc
quenching is required to reproduce the measured lifetime; the correlations
induced by the core-polarization produce
about two-thirds of the required reduction themselves.

Turning now to the WIMP structure functions in fig. 2, we see that all three
are strongly quenched from their single-particle values.  The strongest
quenching is in $S_{00}(q)$, which is reduced to 25\% and 20\% of the
single-particle value for the two residual interactions considered.  Our
preferred choice, the solid line (corresponding to interaction II), is the
lowest in all three diagrams.  The solid lines are accurately reproduced by the
following fourth-order polynomials in $y=(bq/2)^2$:
\begin{eqnarray}
\label{e:f2}
S_{00} & = & 0.0094999 - 0.0619718 \,y + 0.162844 \,y^2  - 0.194282 \,y^3 +
0.0891054 \,y^4
\\ \nonumber
S_{11} & = & 0.0298127 - 0.2176360 \,y + 0.623646 \,y^2 - 0.814418 \,y^3 +
0.4050270 \,y^4  \\ \nonumber
S_{01} & = & 0.0332044 - 0.2319430 \,y + 0.638528 \,y^2 - 0.798523 \,y^3 +
0.3809750 \,y^4~.
\end{eqnarray}
We repeat that these structure functions completely determine spin-dependent
cross sections for any heavy Majorana particle.

\section{Discussion}
\label{s:d}

What are the prospects for mica as a WIMP detector?  At present the limits that
can be set are several orders of magnitude higher than the
expected cross sections for
neutralinos.  But by analyzing more mica, the experiments may reach the
necessary level of sensitivity within a few years.  Because the efficiency with
which each element in mica can be detected when it recoils is different, we
cannot easily present
an overall cross section for a representative neutralino, as is the
common practice in the literature.  But we note that the calculated
efficiencies\cite{r:Ifft} begin to increase at a recoil energy of about 20 keV,
corresponding to a $q^2$ of about $10^{-3} {\rm GeV}^2$ in Al, and about $1.6
\times 10^{-3} {\rm GeV}^2$ in K.
If the 20 keV were a detection threshold, then as
the curves in figs. 1 and 2 show, the number of events below threshold
would be a small fraction of the total.

Whatever the precise characteristics of the detector, the expected number of
events can now be calculated
by folding the efficiencies together with the cross sections presented here and
with the flux of WIMPs, which, unfortunately, is still poorly constrained.
But continuing work with the mica, with complementary detectors, and
with telescopes should make it possible to
rule in or out most hypothetical WIMPs within the next decade.

J.E. is supported in part by the U.S.\ Department of Energy under grant
DE-FG05-94ER40827.  M.T.R. is supported by the Weingart Foundation. Part of
the computing was carried out at Lawrence Livermore National Laboratory,
operated under the auspices of the U.S.\ Department of
Energy under grant W-7405-ENG-48.
Oak Ridge National Laboratory is managed for the U.S.\ Department of Energy
by Martin Marietta Energy Systems, Inc. under contract No.
DE--AC05--84OR21400. Theoretical nuclear physics research at the University of
Tennessee is supported by the U.S.\ Department of Energy through contract No.
DE-FG05-93ER40770.

We owe thanks to Buford Price, David Resler, Dan Snowden-Ifft, and
Petr Vogel for useful discussions.

\begin{figure}
\caption{The three $^{27}$Al structure functions
$S_{00}$ (top), $S_{11}$ (middle), and
$S_{01}$ (bottom) as a function of momentum-transfer squared.  The dotted lines
are the single-particle functions, the dashed lines the full functions with
harmonic oscillator single-particle wave functions, and the solid lines the
full functions with Woods-Saxon single-particle wave functions.}
\label{f:1}
\end{figure}

\begin{figure}
\caption{The three $^{39}$K structure functions
$S_{00}$ (top), $S_{11}$ (middle), and
$S_{01}$ (bottom) as a function of momentum-transfer squared.  The dotted lines
are the single-particle functions, the dashed lines the full functions
obtained from the modified Bonn interaction, and the solid lines the full
functions obtained from the Paris-based G-matrix (see text). }
\label{f:2}
\end{figure}

\begin{table}
\caption{The value of (the z-projection of) the
nuclear spin and orbital-angular-momentum matrix elements for
$^{27}$Al.
\label{t:1}}
\begin{tabular}{lcccc}
\tableline
& $\langle {\bf S}_p \rangle$ & $\langle {\bf S}_n \rangle$ &
$\langle {\bf L}_p \rangle$ & $\langle {\bf L}_n \rangle$ \\
 ISPSM & 0.5 & 0 & 2.0 & 0 \\
 OGM & 0.25 & 0 & 2.25 & 0  \\
 EOGM ($g_A/g_V = 1.0$)  & 0.333 & 0.043 & --- & ---  \\
 EOGM ($g_A/g_V = 1.25$)  & 0.304 & 0.073 & --- & ---  \\
 Shell Model & 0.3430 & 0.0296 & 1.7812 & 0.3461  \\
\end{tabular}
\end{table}

\begin{table}
\caption{The value of (the z-projection of) the
nuclear spin and orbital-angular-momentum matrix elements for
$^{39}$K.
\label{t:2}}
\begin{tabular}{lcccc}
\tableline
& $\langle {\bf S}_p \rangle$ & $\langle {\bf S}_n \rangle$ &
$\langle {\bf L}_p \rangle$ & $\langle {\bf L}_n \rangle$ \\
 ISPSM & -0.3 & 0 & 1.8 & 0 \\
 OGM & --- & 0 & --- & 0  \\
 EOGM ($g_A/g_V = 1.0$)  & -0.196 & 0.055 & --- & ---  \\
 EOGM ($g_A/g_V = 1.25$)  & -0.171 & 0.030 & --- & ---  \\
 Force I & -0.197 & 0.051 & 1.339 & 0.308  \\
 Force II& -0.184 & 0.054 & 1.068 & 0.564 \\
\end{tabular}
\end{table}

\begin{table}
\caption{Corrections to the single-particle $(sp)$ values of the
isoscalar ($\mu_{S}$) and isovector ($\mu_{V}$) magnetic moments and
diagonal Gamow-Teller beta-decay matrix element ($M(GT)$),
for the mass $A=39$ system from a Bonn-based, weak-tensor force,
$I$, and a Paris-based, strong-tensor force, $II$.  The corrections are due
primarily to  second-order core-polarization $(CP)$ and
meson-exchange currents $(MEC)$. See text for further explanation.
\label{t:3}}
\begin{tabular}{lcrrrrr}
\tableline
 & Force & $sp$~~ & $CP$~ & MEC & Sum & Expt  \\
$\mu_{S}$ & I & 0.636 & 0.066 & 0.004 & 0.706 & 0.707 \\
          & II & 0.636 & 0.064 & 0.008 & 0.708 & 0.707 \\
$\mu_{V}$ & I & $-0.512$ & $-0.138$ & 0.364 & $-0.286$ & $-0.31$5 \\
          & II & $-0.512$ & $-0.378$ & 0.369 & $-0.521$ & $-0.315$
\\
$M(GT)$   & I & $-0.976$ & 0.176 & $-0.009$ & $-0.809$ & $-0.647$ \\
          & II & $-0.976$ & 0.205 & $-0.016$ & $-0.787$ & $-0.647$ \\
\end{tabular}
\end{table}

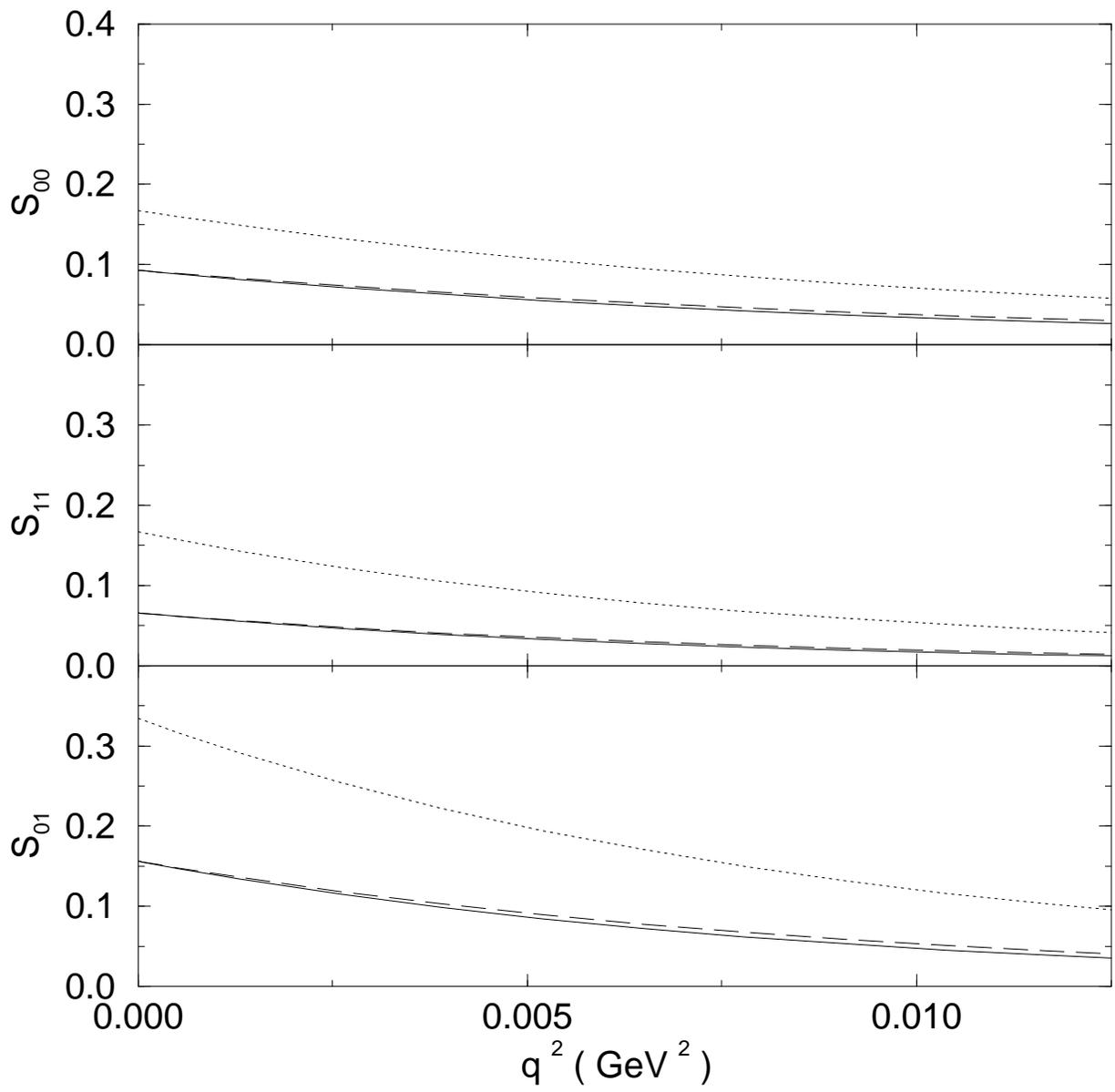
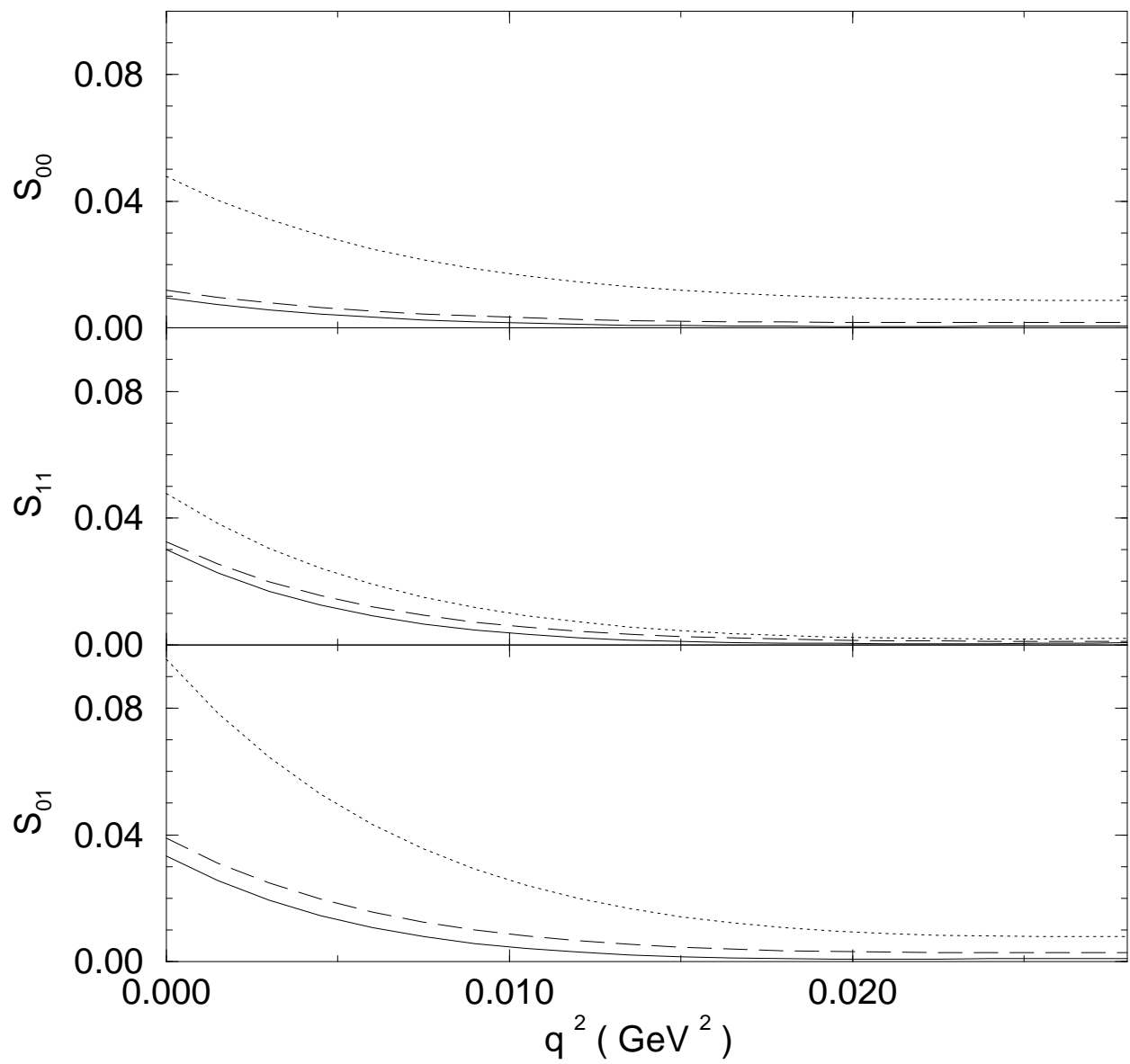
\end{document}